\documentclass[aps,twocolumn,showpacs]{revtex4}
\usepackage{graphicx}

\begin{document}

\author{I.M. Sokolov} 
\affiliation{Institut f{\"u}r Physik, Humboldt-Universit{\"a}t zu Berlin,
Newtonstra{\ss}e 15, D-12489 Berlin, Germany}
\author{A.V. Chechkin}
\affiliation{Institute for Theoretical Physics, N SC KIPT Akademicheskaya
st. 1, 61108 Kharkov, Ukraine}
\author{J. Klafter}
\affiliation{School of Chemistry, Sackler Faculty of Exact Sciences, Tel
Aviv University, Tel Aviv 69978, Israel}

\title{Fractional Diffusion Equation for a Power-Law-Truncated L\'{e}vy Process}

\date{\today}

\begin{abstract}
Truncated L\'{e}vy flights are stochastic processes which display a
crossover from a heavy-tailed L\'{e}vy behavior to a faster decaying
probability distribution function (pdf). Putting less weight on long flights
overcomes the divergence of the L\'{e}vy distribution second moment. We
introduce a fractional generalization of the diffusion equation, whose
solution defines a process in which a L\'{e}vy flight of exponent $\alpha \ $%
is truncated by a power-law of exponent $5-\alpha $. A closed form for the
characteristic function of the process is derived. The pdf of the
displacement slowly converges to a Gaussian in its central part showing
however a power law far tail. Possible applications are discussed.
\end{abstract}


\maketitle

Since L\'{e}vy flights have been introduced into statistical physics, it has
become clear that special attention must be given to the fact that due to
their heavy tails they are characterized by diverging moments. A few
approaches have been suggested to overcome this divergence. These include
the introduction of the concept of L\'{e}vy walks \cite{SWK}, confining
L\'{e}vy flights by external potentials \cite{ChechKla}, and introduction of
truncation procedures \cite{ManSte,Koponen}. Each of the approaches
represents a different physical situation, but they all made it possible for
L\'{e}vy processes to be applicable in a variety of areas ranging from
Hamiltonian dynamics \cite{Hamilto} to spectral diffusion in single molecule
spectroscopy \cite{SingMol}, from bacterial motion \cite{Bacteria} to the
albatross flights \cite{Albatros} and also in analysis of economical data
\cite{R1,R2}. Here we concentrate on the truncation of the flights.

In many cases the L\'{e}vy-flight behavior corresponds to intermediate
asymptotics. At very large values of the variable some cutoff enters, so
that the moments exist. Truncated L\'{e}vy flights, a process showing a slow
convergence to a Gaussian, were introduced by Mantegna and Stanley \cite
{ManSte} and have been since used especially in econophysics, see Refs. 
\cite{R1,R2}. The truncated L\'{e}vy flight is a Markovian jump process,
with the length of jumps showing a power-law behavior up to some large scale. 
At larger scales the power-law behavior crosses over to a faster
decay, so that the second moment of the jump lengths exists. In this case
the central limit theorem applies, so that at very long times the
distribution of displacements converges to a Gaussian; this convergence
however may be extremely slow. The original work concentrated on numerical
simulations of the process which assumed a $\theta $-function cutoff.
Koponen \cite{Koponen} slightly changed the model by replacing the $\theta $%
-function cutoff by an exponential one and obtained a useful analytical
representation for the model. However, further investigations have shown
that the models with sharp ($\theta $-function or exponential) cutoffs
predicting a Gaussian or an exponential tail of the pdf are not always
appropriate \cite{R2}.

In mathematical physics, it is often convenient to have a deterministic
equation for the pdf of a process, an analogue of the diffusion or
Fokker-Planck equation (FPE), to be solved under given initial and boundary
conditions. For the case of anomalous transport, fractional generalizations
of such equations may be relevant \cite{phys2day}. However, these classes of
equations are valid only for processes showing exact scaling in the
force-free limit. The truncated L\'{e}vy flights are not such a process. As
mentioned, in the course of time the process displays a crossover between
the two regimes: at shorter times the characteristics of this random process
behave as those of a L\'{e}vy flight, while at long times they are close to
ones for normal diffusion. Processes showing a crossover can often be
described by equations containing derivatives of different order in the same
variable. A known example here is the telegrapher's equation with the second
and the first-order temporal derivatives, describing the crossover from
ballistic transport to the diffusion behavior. However, fractional
generalizations of the telegrapher's equation describe L\'{e}vy-walk-like
processes \cite{LeWa} and not truncated L\'{e}vy flights.

The equation we propose for truncated L\'{e}vy flights with the power-law
cutoff has the following form:

\begin{equation}
\left( 1-C_{\alpha }\frac{\partial ^{2-\alpha }}{\partial \left| x\right|
^{2-\alpha }}\right) \frac{\partial p(x,t)}{\partial t}=D\frac{\partial
^{2}p(x,t)}{\partial x^{2}},  \label{MainEq}
\end{equation}
where $D$ is the diffusion coefficient governing the long-time asymptotic
behavior, and the scale factor $C_{\alpha }=D/K_{\alpha }$ is a coefficient
governing the intermediate-time L\'{e}vy-like one. The dimension of $%
C_{\alpha }$ is $\left[ C_{\alpha }\right] =[L^{2-\alpha }]$. In Eq.(\ref
{MainEq}) $\frac{\partial ^{\alpha }}{\partial \left| x\right| ^{\alpha }}$
denotes the symmetric Riesz-Weyl operator \cite{Samko}, which can be
expressed through 
\begin{equation}
\frac{d^{\beta }}{d\left| x\right| ^{\beta }}f(x)= -\frac{1}{2\cos (\pi
\beta /2)}\left[ _{-\infty }D_{x}^{\beta }+~_{x}D_{-\infty }^{\beta }\right]
\label{Deriv}
\end{equation}
for $0<\beta <2$, $\beta \neq 1$ and 
\begin{equation}
\frac{d^{\beta }}{d\left| x\right| ^{\beta }}f(x)= -\frac{d}{dx}\hat{H}f(x)
\label{Deriv1}
\end{equation}
for $\beta =1$, where $~_{-\infty }D_{x}^{\beta }$ and $~_{x}D_{-\infty
}^{\beta }$ are the corresponding Riemann-Liouville operators, and $\hat{H}$
denotes the Hilbert transform 
\begin{equation}
\hat{H}\phi =\frac{1}{\pi }\int_{-\infty }^{\infty }\frac{\phi (\xi )d\xi }{%
x-\xi }.
\end{equation}
The operator defined by Eqs. (\ref{Deriv}) and (\ref{Deriv1}) is a
fractional generalization of the \emph{second} derivative: in
Fourier-representation $\frac{\partial ^{\beta }}{\partial \left| x\right|
^{\beta }}\phi (x)$ simply corresponds to $-\left| k\right| ^{\beta }\phi
(k) $, which for $\beta =2$ gives us a known form $-k^{2}\phi (k)$. This is
the reason to include the ''minus'' sign in Eq.(\ref{Deriv}). Note however
that $\frac{\partial ^{0}}{\partial \left| x\right| ^{0}}\phi (x)$
corresponds to $-\phi (k)$.

Eq.(\ref{MainEq}) is a special case of the distributed-order diffusion
equation: 
\begin{equation}
-\int_{0}^{2}d\alpha ^{\prime }f(\alpha ^{\prime })C_{\alpha ^{\prime }}%
\frac{\partial ^{2-\alpha ^{\prime }}}{\partial \left| x\right| ^{2-\alpha
^{\prime }}}\frac{\partial p(x,t)}{\partial t}=D\frac{\partial ^{2}p(x,t)}{%
\partial x^{2}}  \label{DOFPE}
\end{equation}
corresponding to a weight function $f(\alpha ^{\prime })=\delta (2-\alpha
^{\prime })+\delta (\alpha -\alpha ^{\prime })$ which describes a crossover
between $\alpha =2$ (normal diffusion) and to $\alpha <2$ (L\'{e}vy-like
superdiffusion). It can be shown, that contrary to the equations used in 
\cite{CGS,SCK} which describe processes getting more and more anomalous in
the course of time (retarding subdiffusion or accelerating superdiffusion), 
equation (\ref{DOFPE}) describes processes getting less
anomalous, such as in our case, tending to normal diffusion. General
properties of this equation will be considered elsewhere.

Using the Fourier-representation of the Riesz-Weyl derivative in Eq.(\ref
{MainEq}) we get for the characteristic function of the distribution: 
\begin{equation}
\left( 1+C_{\alpha }\left| k\right| ^{2-\alpha }\right) \frac{\partial f(k,t)%
}{\partial t}=-Dk^{2}f(k,t).
\end{equation}
The Green's function of this equation, corresponding to the initial
condition $f(k,0)=1$ (i.e. $p(x,0)=\delta (x)$) then reads: 
\begin{equation}
f(k,t)=\exp \left( -\frac{Dk^{2}}{1+C_{\alpha }\left| k\right| ^{2-\alpha }}%
t\right) .  \label{Chaf}
\end{equation}
We postpone the proof of the fact that this is indeed a characteristic
function of some probability distribution $p(x,t)$ until later on and
discuss first the main properties of such a solution.

The function $f(k,t)$ is differentiable twice for each $\alpha $; its second
derivative $\left. f^{\prime \prime }(k,t)\right| _{k=0}=-2Dt$, so that the
second moment of the distribution evolves in a diffusive manner: $%
\left\langle x^{2}(t)\right\rangle =2Dt$, as in normal diffusion. However,
in the intermediate domain of $x$ the distribution shows the behavior
typical for L\'{e}vy flights; namely for $k$ large enough, i.e. for $%
C_{\alpha }\left| k\right| ^{2-\alpha }\gg 1$, the characteristic function
has the form 
\begin{equation}
f(k,t)=\exp \left( -\frac{D}{C_{\alpha }}\left| k\right| ^{\alpha }t\right) ,
\label{Middle}
\end{equation}
i.e. corresponds to the characteristic function of the L\'{e}vy
distribution. Assuming $\alpha <2$ we get the following expansion for $%
f(k,t) $ near $k=0$: 
\begin{equation}
f(k,t)\simeq 1-Dtk^{2}+DC_{\alpha }t\left| k\right| ^{4-\alpha }+...
\label{Chaf2}
\end{equation}
From this expression it is evident that $f(k,t)$ always lacks the fourth
derivative at $k=0$ (for $1<\alpha <2$ it even lacks the third derivative),
which means that the fourth moment of the corresponding distribution
diverges. The absence of higher moments of the distribution explains the
particular nature of the truncation implied by our model: The L\'{e}vy
distribution is truncated by a power law with a power between 3 and 5. Thus,
for all $0<\alpha <2$ the corresponding distributions have a finite
second moment and, according to the central limit theorem (slowly!) converge
to a Gaussian. For the case $0<\alpha <1$ (when $\left\langle \left|
x^{3}\right| \right\rangle <\infty $) the speed of this convergence is given
by the Berry-Esseen theorem, as noted in Ref.\cite{Shles}. The convergence
criteria for $1<\alpha <2$ can be obtained using theorems of Ch. XVI of Ref. 
\cite{Feller}.

This transition from the initial L\'{e}vy-like distribution to a Gaussian is
illustrated in Fig.1, obtained by a numerical inverse Fourier-transform of
the characteristic function, Eq.(\ref{Chaf}). Here the case $\alpha =1,D=C=1$
is shown. To put the functions for $t=0.001$ and for $t=1000$ on the same
plot we rescale them in such a way that the characteristic width of the
distribution $W(t)$ (defined by $\int_{0}^{W(t)}p(x,t)dx=\frac{1}{4})$ is
the same. The behavior of the pdf to be at the origin $p(0,t)$ as a function
of $t$ is shown on the double logarithmic scales in Fig.2. Note the
crossover from the initially fast decay $p(0,t)\propto t^{-1/\alpha }$
(L\'{e}vy superdiffusion) to the final form $p(0,t)\propto t^{-1/2}$ typical
for diffusion.

\begin{figure}[tbp]
\includegraphics{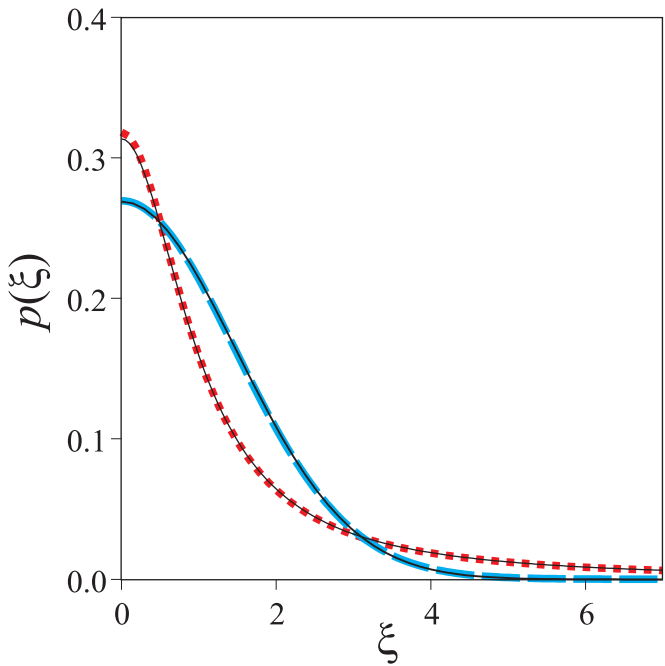}
\caption{The rescaled pdf $p(\xi )=W(t)p(x,t)$ is shown for $t=0.001$
(dotted lime) and for $t=1000$ (dashed line) as a function of a rescaled
displacement $\xi =x/W(t)$. The corresponding thin lines denote the limiting
Cauchy and Gaussian distributions under the same rescaling.}
\end{figure}

\begin{figure}[tbp]
\includegraphics{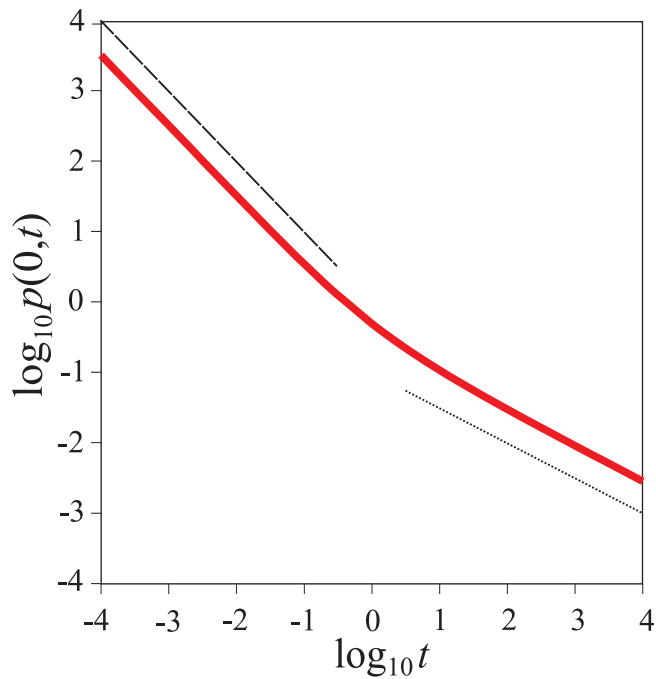}
\caption{Shown is the pdf to be at the origin $p(0,t)$ as a function
of time, see text for details. Note the double logarithmic scales. The
dashed line has the slope $-1$, and corresponds to the superdiffusive decay;
the dotted line has the slope of $-1/2$, as in the case of normal diffusion.}
\end{figure}

The asymptotics of the pdf at large $x$ is determined by the first
non-analytical term in the expansion, Eq.(\ref{Chaf2}), i.e. by $DC_{\alpha
}t\left| k\right| ^{4-\alpha }$. By making the inverse Fourier
transformation of this term and using the Abel method of summation of
improper integral, we get

\begin{equation}
p(x,t)\simeq \frac{\Gamma (5-\alpha )\sin (\pi \alpha /2)}{\pi }\frac{%
DC_{\alpha }t}{x^{5-\alpha }},\qquad x\rightarrow \infty ,  \label{FarTail}
\end{equation}
where we use that $\int_{0}^{\infty }d\xi ~\xi ^{4-\alpha }\cos \xi =\sin
(\pi \alpha /2)\Gamma (5-\alpha )$. Thus, in our case the
L\'{e}vy-distribution is truncated not by a $\theta $- or exponential
function, but by a steeper power-law, with a power $\beta =5-\alpha $.

For example a L\'{e}vy flight truncated by another, faster decaying
power-law, is a much better model for the behavior of commodity prices.
Thus, the discussion in Ref.\cite{R2} shows that the cumulative distribution
function of cotton prices may correspond to a power-law behavior of $%
1-F(x)=\int_{x}^{\infty }p(x)dx\propto x^{-\alpha }$ with the power $\alpha
=1.7$ in its middle part and with the far tail decaying as a power-law $%
1-F(x)\propto x^{-\beta }$ with $\beta \approx 3$. Thus, our equation (which
is definitely the simplest form of the equation for truncated L\'{e}vy
flights) adequately describes this very interesting case giving $\beta =3.3$%
. It is highly probable that fractional equations of the type considered
here might be a valuable tool in economic research.

Let us now prove that the solution $p(x,t)$ is a pdf, i.e. a non-negative
normalized function of $x$ for any $t$. The normalization is trivial and
follows from the fact that $\hat{f}(0,t)=1$ for all $t$. Let us now prove
the non-negativity of the solution.

We start from defining a function $G(u,t)$, $u>0$, such that its Laplace
transform in variable $u$ is 
\begin{equation}
\tilde{G}(s,t)=\int_{0}^{\infty }du\ e^{-su}G(u,t)=\exp \left( -\frac{s}{%
1+A_{\alpha }s^{1-\alpha /2}}t\right)  \label{fs}
\end{equation}
with $A_{\alpha }=C_{\alpha }/D^{1-\alpha /2}$. By comparing Eqs.(\ref{fs})
and (\ref{Chaf}), one sees that the characteristic function $\hat{f}(k,t)$
can be rewritten in the following form:

\begin{equation}
\hat{f}(k,t)=\int_{0}^{\infty }e^{-uDk^{2}}G(u,t)du.  \label{fk}
\end{equation}
Indeed, the transition from Eq.(\ref{Chaf}) to Eqs.(\ref{fk}) is nothing
else but the change of variable $s\rightarrow Dk^{2}$ in Eq.(\ref{fs}). It
is clear that $\tilde{G}(0,t)=1$ and, moreover, as we proceed to show, $%
\tilde{G}(s,t)$ is completely monotonic, i.e. it is non-negative, and the
signs of its derivatives alternate. Then, according to Bernstein's theorem 
\cite{Feller}, $\tilde{G}(s,t)$ is a Laplace-transform of some probability
density. Now, we can perform the inverse Fourier-transform in the Eq.(\ref
{fk}) and get 
\begin{equation}
p(x,t)=\int_{0}^{\infty }\frac{1}{\sqrt{4\pi Du}}\exp \left( -\frac{x^{2}}{%
4Du}\right) G(u,t)du  \label{Subord1}
\end{equation}
which is a nonnegative function (since the integrand is a product of two
non-negative functions). Eq.(\ref{Subord1}) provides a subordination
transformation: the truncated L\'{e}vy flights can be considered as a
process subordinated to a Wiener process under the operational time given by
the function $G(u,t)$ \cite{Sok}. The small-$u$ behavior of this function
(corresponding to the large-$s$ one of its Laplace-transform, $\tilde{G}%
(s,t)=\exp \left( -A_{\alpha }^{-1}s^{\alpha /2}t\right) $) is approximately
a one-sided (extreme) L\'{e}vy law of index $\alpha /2$. However, at large $%
u $ this L\'{e}vy law is truncated.

We now give a proof that the function $\tilde{G}(s,t)=\exp \left( -\frac{s}{%
1+A_{\alpha }s^{1-\alpha /2}}t\right) $ is a completely monotonic function
in variable $s$. This function has a form $\exp (-\psi (s))$ and therefore
is completely monotonic if the function $\psi (s)$ is positive and possesses
a completely monotonic derivative. In our case $\psi (s)=s/(1+As^{b})$ with $%
b=1-\alpha /2,$ $0<b<1$, $A>0$. Its derivative is given by 
\begin{equation}
\psi ^{\prime }(s)=\frac{1}{1+As^{b}}\left[ 1-b\frac{As^{b}}{1+As^{b}}%
\right] .
\end{equation}
This function is a product of two functions. The first one is completely
monotonic since it has a form $g(h(s))$ with $g(y)=1/(1+y)$ being completely
monotonic and with $h(s)=As^{b}$ being a positive function with a completely
monotonic derivative. The second function has the same form, now with $%
g(y)=1-by/(1+y)$. This function $g(y)$ is positive for all $y>0$, and its
derivatives read $g^{(n)}(y)=(-1)^{n}bn!(1+y)^{-n-1}$.

The subordination property also sheds light on the possible nature of
truncated L\'{e}vy distributions in economic processes. The truncated
L\'{e}vy process can be interpreted as a simple random walk with a finite
variance. However, the number of steps of the random walk (the number of
transactions) per unit time is not fixed, but fluctuates strongly. The
implications to economics of such models were considered in \cite{Clark}. In
our case the distribution function of the number of steps has itself a form
of a truncated one-sided L\'{e}vy law.

Let us summarize our findings. We proposed a fractional generalization of a
diffusion equation which describes power-law truncated L\'{e}vy flights, a
random process showing a slow convergence to a Gaussian. We show that the
solution of this equation is a pdf and give numerical results for the case $%
\alpha =1$. Moreover, we argue that the truncated L\'{e}vy flights can be
represented as a random process subordinated to a Wiener process, which
might be helpful in econophysical applications. We end by noting that the
equation discussed is a special case of distributed-order fractional
diffusion equations. Modifications of our equation should be able to
describe other types of truncation; however, the equation discussed here is
definitely the simplest one.

IMS acknowledges partial financial support by the Fonds der Chemichen
Industrie. AVC and JK acknowledge the support within the INTAS 00-0847
project.

\end{document}